\documentclass{iaus}
\usepackage{graphicx}
\usepackage{natbib}

\title[Resonance capture] 
{Hamiltonian model of capture into mean motion resonance}

\author[Alexander J. Mustill \& Mark C. Wyatt]   
{Alexander J. Mustill$^1$
 \and Mark C. Wyatt$^1$}

\affiliation{$^1$Institute of Astronomy, University of Cambridge, \\ Madingley Road,
CB3 0HA, Cambridge, UK \\ email: {\tt ajm233@ast.cam.ac.uk}, {\tt wyatt@ast.cam.ac.uk} \\[\affilskip]}

\pubyear{2010}
\volume{276}  
\pagerange{000--000}
\jname{The astrophysics of planetary systems: formation, structure and dynamical evolution}
\editors{A. Sozzetti, M.G. Lattanzi \& A.P. Boss, eds.}
\begin{document}

\maketitle

\begin{abstract}
Mean motion resonances are a common feature of both our own Solar System and of extrasolar planetary systems. Bodies can be trapped in resonance when their orbital semi-major axes change, for instance when they migrate through a protoplanetary disc. We use a Hamiltonian model to thoroughly investigate the capture behaviour for first and second order resonances. Using this method,  all resonances of the same order can be described by one equation, with applications to specific resonances by appropriate scaling. We focus on the limit where one body is a massless test particle and the other a massive planet. We quantify how the the probability of capture into a resonance depends on the relative migration rate of the planet and particle, and the particle's eccentricity. Resonant capture fails for high migration rates, and has decreasing probability for higher eccentricities, although for certain migration rates, capture probability peaks at a finite eccentricity. We also calculate libration amplitudes and the offset of the libration centres for captured particles, and the change in eccentricity if capture does not occur. Libration amplitudes are higher for larger initial eccentricity. The model allows for a complete description of a particle's behaviour as it successively encounters several resonances. The model is applicable to many scenarios, including (i) Planet migration through gas discs trapping other planets or planetesimals in resonances; (ii) Planet migration through a debris disc; (iii) Dust migration through PR drag. The Hamiltonian model will allow quick interpretation of the resonant properties of extrasolar planets and Kuiper Belt Objects, and will allow synthetic images of debris disc structures to be quickly generated, which will be useful for predicting and interpreting disc images made with ALMA, Darwin/TPF or similar missions. Full details can be found in \cite{2010submitted}.
\end{abstract}

\firstsection 
\section{Introduction}

Mean motion resonances (MMRs) occur when two objects' orbital periods are close to a ratio of two integers, and a particular combination of orbital angles, the resonant argument, is librating. Examples in the Solar System include Neptune and Pluto (3:2 resonance) and the inner Galilean moons of Jupiter (4:2:1 Laplace resonance). There are also now numerous examples of suspected or confirmed MMRs in extrasolar planetary systems (e.g., GJ~876~b and c in a 2:1 resonance, \citealt{2001ApJ...551L.109L}).

Mean motion resonances also occur between planets and small dust particles, as seen in the Earth's resonant dust ring~\citep{1994Natur.369..719D}. Some extrasolar debris discs, such as Vega, show evidence of non-axisymmetric clumps \citep{1998Natur.392..788H,2002ApJ...569L.115W}, and several authors have attempted to model these as arising from a planet's resonant perturbations \citep[e.g.,][]{2003ApJ...588.1110K,2003ApJ...598.1321W}.

Resonances are common because convergent migration between orbiting bodies can cause them to become captured. There are many mechanisms by which such a semi-major axis change can be driven. Early work looked at the tidal evolution of satellite orbits \citep{1965MNRAS.130..159G}. In a protoplanetary disc, planets can migrate by tidal interaction with the gas disc \citep[see][for a recent review]{2009AREPS..37..321C}, and small planetesimals by aerodynamic drag \citep{1977MNRAS.180...57W}. In a gas-depleted debris disc, planets can migrate by gravitational scattering of planetesimals \citep{2009Icar..199..197K}. Interplanetary dust drifts towards the Sun under the influence of Poynting-Robertson (PR) drag \citep{1979Icar...40....1B}.

Resonance capture has been studied by several authors. The regime of adiabatic migration, where the migration timescale is much longer than the resonant argument's libration timescale, has been studied extensively analytically using a Hamiltonian model \citep[e.g.,][]{1982CeMec..27....3H,1984CeMec..32..127B}. Rapid migration was studied using full N-body models by \citet[henceforth W03]{2003ApJ...598.1321W} for the case of a planet migrating into a planetesimal disc, and using the Hamiltonian model by \citet[henceforth Q06]{2006MNRAS.365.1367Q} for general migration scenarios. Q06 obtained capture probability as a function of migration rate and eccentricity. We have extended this work, and using the Hamiltonian model we calculate capture probabilities, libration amplitudes and offsets for particles that are captured, and eccentricity jumps for those that pass through the resonance without capture.

The Hamiltonian model we use has several advantages over N-body simulations: (1) it allows some results to be derived analytically; (2) it is faster to integrate numerically than the 3-body problem; (3) all resonances of the same order reduce to a Hamiltonian of the same form, with fewer free parameters than the three-body problem. Once a suite of numerical integrations of the Hamiltonian model is performed, it can be applied to any system, without the need for running a different N-body integration every time the system parameters are changed.

\begin{figure}
\begin{center}
 \includegraphics[width=.45\textwidth]{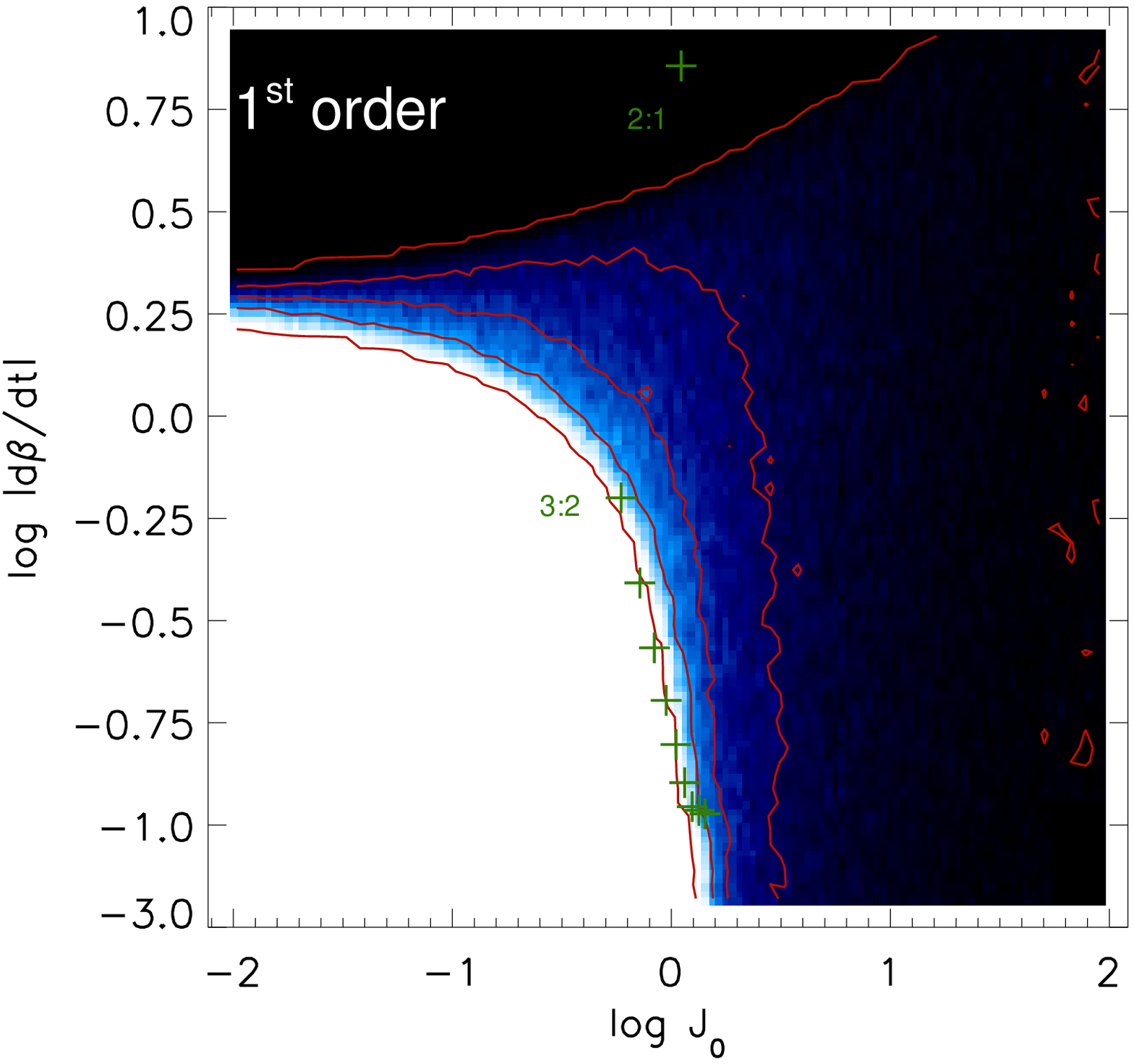}
\includegraphics[width=.45\textwidth]{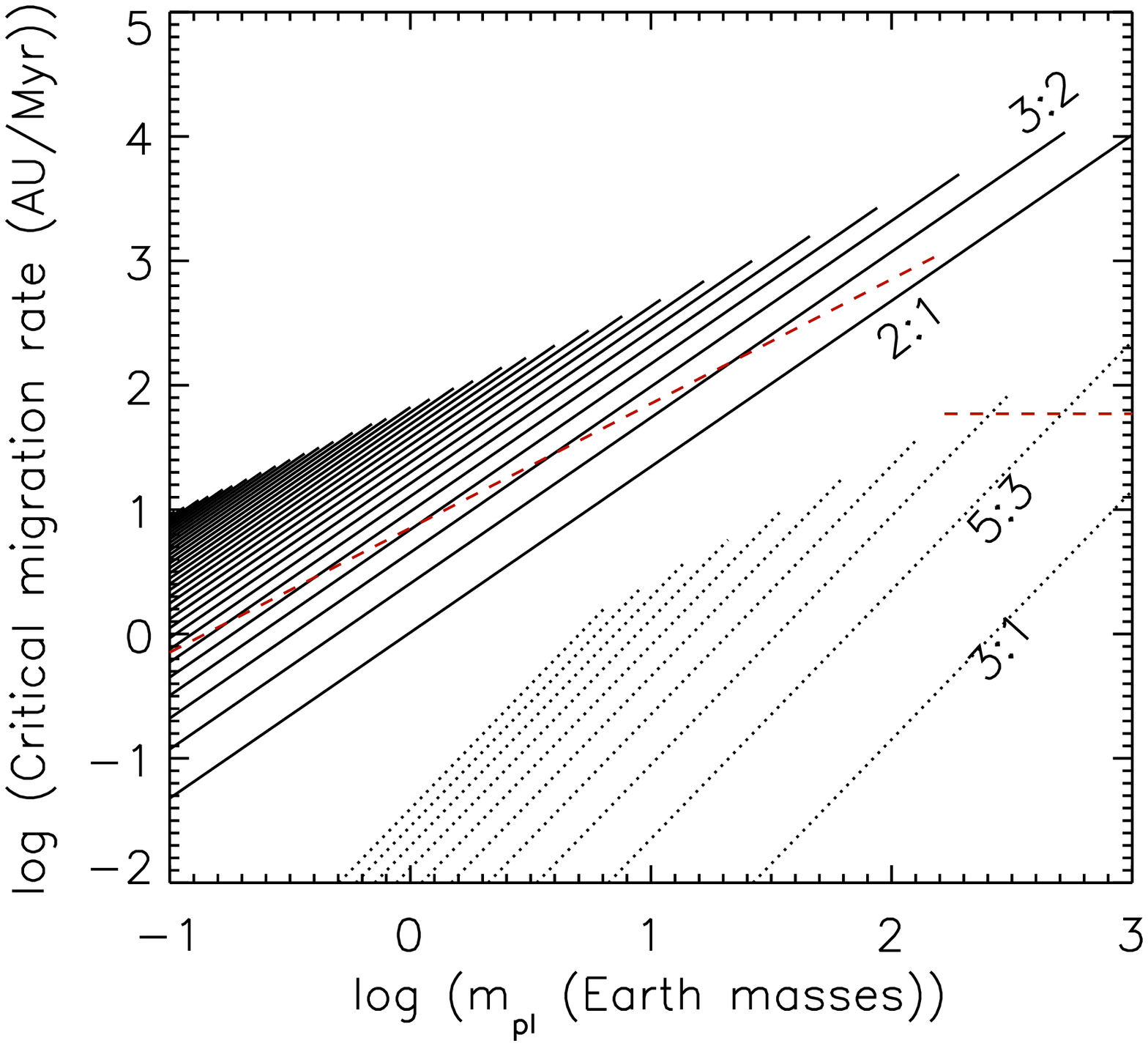}
 \caption{\textbf{Left: }Capture probability for first order resonances, as a function of rescaled eccentricity $J_0$ and migration rate $\dot\beta$. Colour scale and contours show capture probability, from white (100\%) to black (0\%). The crosses show the rescaled eccentricity and migration rate corresponding to a particle with Keplerian eccentricity 0.01, migrating at 1\,AU\,Myr$^{-1}$ into exterior resonances with an Earth-mass planet orbiting a Solar-mass star at 1\,AU. The 2:1 and 3:2 resonances are labelled. \textbf{Right: }Critical migration rate at 1\,AU for capture of low-eccentricity particles beign trapped into resonances with a migrating planet.  At migration rates faster than this, capture is impossible at low eccentricity. The critical migration rates for first order resonances are shown as solid lines. We show typical migration rates for planets embedded in gas discs at 1\,AU (dashed lines). Mass-dependent Type I migration occurs for low mass planets and mass-independent Type II migration for high mass planets.}
   \label{fig:prob}
\end{center}
\end{figure}

\section{Model}

We work with a Hamiltonian model of mean motion resonance \citep[e.g.,][]{1999ssd..book.....M}. We consider the circular restricted three body problem with a massive planet and a massless test particle orbiting a central star. The test particle orbits either interior or exterior to the planet; the resulting equations of motion are the same, with only a change in the scalings between the physical variables and the rescaled variables of the Hamiltonian model. The Hamiltonian is
\begin{equation}\label{eq: Ham}
\mathcal{H} = J^2 + \beta J - J^{k/2} \cos k\theta,
\end{equation}
where $\theta$ is related to the resonant argument and the conjugate momentum $J$ is proportional to the square of the particle's eccentricity, and $k$ is the order of the resonance. The parameter $\beta$ measures the proximity in semi-major axis to the resonance, and to simulate migration $\beta$ is varied at a constant rate proportional to the physical migration rate. The equations of motion arising from this Hamiltonian were integrated numerically. We varied the initial momentum $J_0$ and the migration rate $\dot\beta$. There are two parameters in the problem: the eccentricity of the particle when it hits the resonance (governed by $J_0$, the initial value of $J$), and the particle's or planet's migration rate (governed by $\dot\beta$). For each point in $(J_0,\dot\beta)$ parameter space we integrated 100 trajectories, with the resonant argument chosen from a uniform distribution over $[0,2\pi)$.

Figure~1 (left) shows capture probabilities for first order resonances as a function of $J_0$ and $\dot\beta$. We see that capture into resonance is guaranteed for small initial eccentricities and migration rates. For low migration rates we are in the well-studied adiabatic regime. For low eccentricities capture is certain since the separatrix forms around the initial trajectory. For high eccentricities the separatrix forms inside the initial orbit and expands to meet it as the migration continues. Capture then is probabilistic, with a probability that decreases as the initial eccentricity increases \citep{1982CeMec..27....3H}. For low eccentricities, with $J_0<1.3$, capture is certain if the migration rate is low and impossible if it is high. In the limit of low eccentricity, the transition occurs at a critical migration rate of $|\dot\beta|\approx 2.1$. Certain capture occurs for low migration rates up to $J_0\approx 1.3$. For higher eccentricities ($J_0>1.3$), capture is always probabilistic, with a capture probability that is not strongly dependent of migration rate; however, if migration rate is too high, then capture is still impossible.

We also found the amplitudes and centres of libration of the resonant argument, and the change in eccentricity if a particle is not captured. In the presence of migration, the centre of libration is offset from the centre in the absence of migration by an amount proportional to the migration rate, almost independent of eccentricity. The amplitude of libration increases slightly with migration rate, and significantly with the particle's initial eccentricity. If a particle is not captured into resonance, its eccentricity is driven down if capture failed due to high eccentricity (in agreement with adiabatic theory; \citealt{1999ssd..book.....M}), up if capture failed due to fast migration but low eccentricity, and either up or down if both eccentricity and migration rate are high.

In \cite{2010submitted} we also investigated second-order resonances.

We validated the model against the N-body integrations of W03. The Hamiltonian model generally shows good agreement with the results of N-body integrations.

\section{Applications}

Using the Hamiltonian model, the result of an encounter with a resonance in many scenarios can be easily determined by rescaling variables and looking up the outcome in the grid of integration results. For example, Figure~1 shows the rescaled eccentricity and migration rate for a particle with Keplerian eccentricity 0.01, migrating at 1\,AU\,Myr$^{-1}$ into exterior resonances with an Earth-mass planet orbiting a Solar-mass star at 1\,AU. We see that it is impossible for the particle to be captured into the 2:1 resonance. If migration continues it is almost certain, however, to be captured into the 3:2 resonance, and if it is not it will be captured into the 4:3 resonance. Because determining the fate of a particle is reduced to looking up the outcome in a data table, the fate can be determined much faster than it can by performing N-body integrations. This makes the model very suitable for generating synthetic images of debris discs with many particles (outward migration of the planet and inward migration of the dust can both be handled in the same way), and for determining the outcome of resonant encounters during planet formation in population synthesis models.

\subsection{Debris discs}

In \cite{2010submitted} we outline how the degree of dynamical excitation of a planetesimal disc can affect the observed structure if a planet migrates into it. More dynamically excited discs will show weaker resonant signatures, since the capture probability is lower and the amplitude of libration higher (see also \citealt{2008A&A...480..551R}). Future work will explore disc structures in detail using this model.

\subsection{Protoplanetary discs}

 In Figure~1 (Right) we show the critical migration rate for capture of particles interior to a migrating planet. Also indicated are typical migration rates. It is straightforward to determine which resonance the planet will capture particles into. For example, a 10 Earth mass planet undergoing Type~I migration will capture bodies into the 4:3 resonance, while a 1000 Earth mass planet undergoing Type~II migration will capture bodies into the 5:3 second-order resonance. Future work will address the issue of how accurate the Hamiltonian model is at describing the dynamics of two massive planets.

\bibliographystyle{apj}
\bibliography{bibliography}

\end{document}